\documentclass[sigconf]{acmart}
\AtBeginDocument{%
  }

\setcopyright{acmcopyright}
\copyrightyear{2023}
\acmYear{2023}
\acmDOI{XXXXXXX.XXXXXXX}

\usepackage{graphicx}
\usepackage{caption}
\usepackage{subcaption}
\usepackage{amsmath}
\usepackage{booktabs}
\usepackage{enumerate}
\usepackage{multirow}
\usepackage{color}

\usepackage[noend]{algpseudocode}
\usepackage{algorithmicx,algorithm}

\newcommand{\vx}{\mathbf{x}}
\newcommand{\vz}{\mathbf{z}}
\newcommand{\vd}{\mathbf{d}}
\newcommand{\vn}{\mathbf{n}}
\newcommand{\mb}{\mathbf}

\acmConference[Preprint paper]{Make sure to enter the correct
  conference title from your rights confirmation emai}{May,
  2023}
  
\acmPrice{15.00}
\acmISBN{978-1-4503-XXXX-X/18/06}
\renewcommand\footnotetextcopyrightpermission[1]{}
\begin{document}

\title{Generative Steganography Diffusion}


\author{Ping Wei}
\email{pwei17@fudan.edu.cn}
\orcid{0000-0002-8852-6618}
\affiliation{%
	\institution{Fudan University}
	\streetaddress{Handan Road 220}
	\city{Shanghai}
	\country{China}}

\author{Qing Zhou}
\email{21110240055@m.fudan.edu.cn}
\affiliation{%
	\institution{Fudan University}
	\streetaddress{Handan Road 220}
	\city{Shanghai}
	\country{China}}

\author{Zichi Wang}
\email{wangzichi@shu.edu.cn}
\affiliation{%
	\institution{Shanghai University}
	\streetaddress{Shangda Road 99}
	\city{Shanghai}
	\country{China}}

\author{Zhenxing Qian}
\authornotemark[1]
\email{zxqian@fudan.edu.cn}
\affiliation{%
	\institution{Fudan University}
	\streetaddress{Handan Road 220}
	\city{Shanghai}
	\country{China}}

\author{Xinpeng Zhang}
\authornote{Xinpeng Zhang and Zhenxing Qian are the corresponding authors.}
\email{zhangxinpeng@fudan.edu.cn}
\affiliation{%
	\institution{Fudan University}
	\streetaddress{Handan Road 220}
	\city{Shanghai}
	\country{China}}
 
\author{Sheng Li}
\email{lisheng@fudan.edu.cn}
\affiliation{%
	\institution{Fudan University}
	\streetaddress{Handan Road 220}
	\city{Shanghai}
	\country{China}}

\renewcommand{\shortauthors}{Ping Wei et al.}

\begin{abstract}

Generative steganography (GS) is an emerging technique that generates stego images directly from secret data. Various GS methods based on GANs or Flow have been developed recently. However, existing GAN-based GS methods cannot completely recover the hidden secret data due to the lack of network invertibility, while Flow-based methods produce poor image quality due to the stringent reversibility restriction in each module. To address this issue, we propose a novel GS scheme called "Generative Steganography Diffusion" (GSD) by devising an invertible diffusion model named "StegoDiffusion". It not only generates realistic stego images but also allows for 100\% recovery of the hidden secret data. The proposed StegoDiffusion model leverages a non-Markov chain with a fast sampling technique to achieve efficient stego image generation. By constructing an ordinary differential equation (ODE) based on the transition probability of the generation process in StegoDiffusion, secret data and stego images can be converted to each other through the approximate solver of ODE -- Euler iteration formula, enabling the use of irreversible but more expressive network structures to achieve model invertibility. Our proposed GSD has the advantages of both reversibility and high performance, significantly outperforming existing GS methods in all metrics.


\end{abstract}




\maketitle

\section{Introduction}
Steganography conceals secret data within cover media for covert communication. It can be applied to various types of media, including digital images \cite{tao2018towards, mandal2022digital}, videos \cite{patel2021study, kunhoth2023video}, audio \cite{wu2020audio}, and text \cite{majeed2021review}. Among these, digital images are the most commonly used cover medium, and the resulting image with hidden data is called the stego image. The purpose of steganography is to prevent unauthorized access to secret data transmission. Steganalysis \cite{subramanian2021image, steganalysis2023image} detects whether a cover or stego image contains hidden data. Traditional steganography can be detected by steganalysis, so generative steganography (GS) is proposed as an alternative method to enhance security.

\begin{figure}[t]
	\centering
	\includegraphics[width=0.8\linewidth]{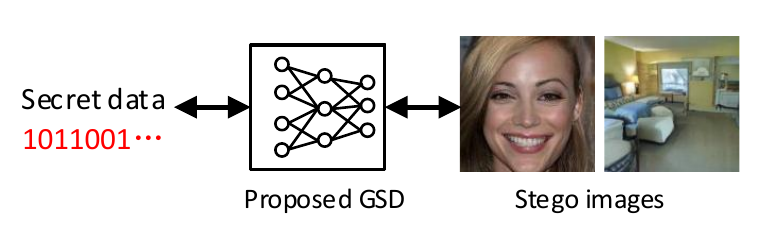} 
	\caption{Generative Steganography with reversibility. Our proposed GSD scheme can convert binary secret data into stego images and then recover  hidden secret data completely.}
	\label{fig:GS}
\end{figure}

Liu \textit{et al}. \cite{liu2017coverless} proposed a generative steganography approach that hides secret data in GANs' class labels, generating different classes of stego images to represent secret data. Similarly, Hu \textit{et al}. \cite{hu2018novel} and Liu \textit{et al}. \cite{ideas} mapped secrets to different latent value ranges of GANs, generating stego images with the image generator, and training an extractor to extract the hidden data. However, their performance is comparatively poor in terms of payload, secret data extraction accuracy, and image quality when compared to traditional carrier-modified-based steganography methods. Later, to increase the payload of secret data, Wei \textit{et al}. \cite{gsn} proposed a generator that directly integrates secret data into convolutional feature maps during image generation, generating high-quality stego images with increased payload, and an extractor is trained to extract hidden data. However, this method has a relatively low data extraction accuracy for high payload capacities. To improve extraction accuracy, Wei \textit{et al}. \cite{gsf} integrated the generator and extractor into one network based on the Flow \cite{glow} model, establishing a bi-directional mapping between stego images and secret data. However, the stego images generated by \cite{gsf} have slightly lower quality than those generated in \cite{gsn}. This is because the invertibility demand in Flow's modules reduces the capability of the network. To obtain higher performance, we propose an innovative GS scheme GSD by constructing a reversible diffusion model. As shown in Fig. \ref{fig:GS}, it can convert binary secret data into quantized stego images and then recover the hidden secret data accurately. Unlike Flow-based methods that use invertible network structures, GSD builds an ordinary differential equation (ODE) to ensure model reversibility. Wherein stego image generation and secret data extraction are completed through the approximate solver of ODE -- Euler iteration formula. This allows for the use of irreversible but expressive networks, which greatly improve model performance.

The diffusion model is currently the most promising image generative method \cite{diffusion_survey,2023diffusion_survey}. Its image quality surpasses that of GAN and demonstrates stunning visual effects \cite{beatgan, GLIDE, dalle-2, stable-diffusion, imagen2022}. The diffusion model consists of a forward diffusion process and a backward image generation process. In the diffusion process, noise is gradually added to the image to obtain a noisy image that satisfies a Gaussian distribution. In the generation process, the noisy image obeying a Gaussian distribution is gradually denoised to obtain a clear natural image. DDPM \cite{ddpm} has utilized the diffusion model to generate high-resolution images, leading to the current trend of the diffusion model. However, the image generation steps of DDPM are too many and take too long. In order to sample more effectively, Song \textit{et al}. propose DDIM \cite{ddim}, which utilizes a non-Markov chain to shorten the sampling steps. OpenAI improves DDPM with a conditional image generation method using classification guidance, surpassing GAN-generated image quality \cite{beatgan}. They also introduce the GLIDE model \cite{GLIDE}, which uses text to guide image generation through the CLIP \cite{CLIP} model or classifier-free guidance. In 2022, OpenAI publishes DALLE-2 \cite{dalle-2}, a two-stage model that generates images by first generating a CLIP image embedding from a text caption and then using a decoder to generate an image. Google later proposes Imagen \cite{imagen2022}, a text-to-image diffusion model that converts input text into text embeddings through a language model and generates high-quality images using robust cascaded diffusion models.

This paper first develops a new diffusion model (designated as StegoDiffusion) for generative steganography, and then proposes a novel GS method based on the proposed StegoDiffusion and DCT/IDCT transformations, called "Generative Steganography Diffusion" (GSD). The data sender can hide secret data in the input latent of the proposed StegoDiffusion by DCT/IDCT transformation and then generate high-quality stego images through the generative process of the model. To retrieve the hidden data, the recipient can input the stego image into StegoDiffusion, which restores the input stego latent using the diffusion process and then extracts the secret data from the DCT coefficients of the recovered latent. The main innovations of our proposed GSD scheme are as follows:

\begin{enumerate}[1)]
\item We first propose an invertible diffusion model for generative steganography that converts secret data into stego images and then recovers the hidden secret data.

\item We construct an ordinary differential equation (ODE) and use its approximate solver (Euler iteration formula) to generate stego images in an invertible manner, allowing the use of irreversible network structures to achieve reversible stego image generation.

\item We explore the use of a non-Markovian chain with a fast sampling technique to speed up the generation of stego images.

\item Our GSD scheme outperforms existing generative steganography methods in all metrics, including payload, extraction accuracy, image quality, and security.


\end{enumerate}

\section{Preliminaries}

\begin{figure}[htb]
	\centering
	\includegraphics[width=0.7 \linewidth]{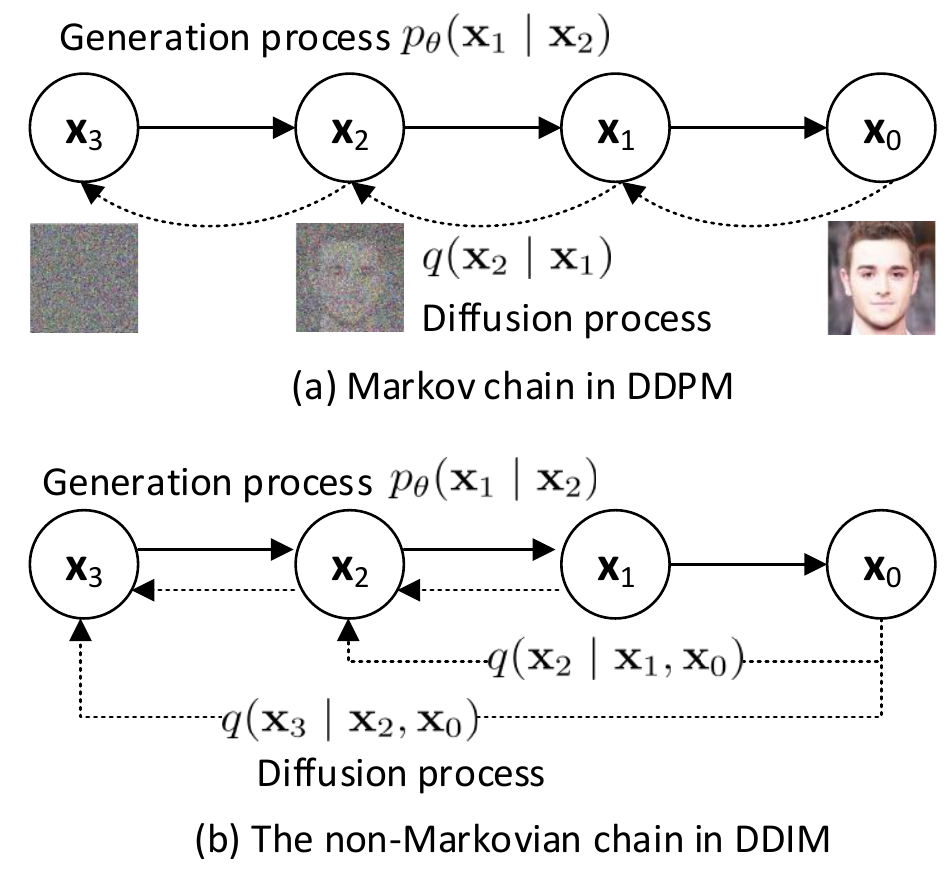} 
	\caption{There are two different diffusion methods. Sub-figure (a) shows the diffusion pattern of the Markov chain in DDPM, where the generation of $\vx_t$ depends only on $\vx_{t-1}$. Sub-figure (b) shows the non-Markovian diffusion method in DDIM, where the generation of $\vx_t$ depends on $\vx_{t-1}$ and $\vx_0$.}
	\label{fig:markov}
\end{figure}

\subsection{DDPM}
DDPM \cite{ddpm} involves two Markov chains: the diffusion process and generation process, as shown in Fig. \ref{fig:markov}a. The diffusion process adds noise to the original image $\mathbf{x}_0$ and generates a noise image $\mathbf{x}_T$ close to the Gaussian distribution through the transition probability $q(\mathbf{x}_t\mid\mathbf{x}_{t-1})$. The generation process, on the other hand, uses a reverse Markov chain $p_{\theta} (\mathbf{x}_{t-1}\mid\mathbf{x}_t)$ to obtain a clear image $\mathbf{x}_0$ from a Gaussian noise image $\mathbf{x}_T$.

The input image distribution is $\mathbf{x}_0 \sim q(\mathbf{x}_0)$, and the diffusion process generates intermediate latents $\mathbf{x}_1, \mathbf{x}_2,..., \mathbf{x}_T$ of the same dimensionality as $\vx_0$
through transition probability $q(\mathbf{x}_t\mid\mathbf{x}_{t-1})$:
\begin{eqnarray}
q\left(\vx_t \mid \vx_{t-1}\right)  = N\left(\vx_t ; \sqrt{\alpha_t} \vx_{t-1}, \beta_t \mathbf{I}\right), 
\end{eqnarray}
where $\alpha_t \in (0,1)$ is a hyper parameter, and $\beta_t = 1- \alpha_t$. Special, when $\Bar{\alpha}_t =\prod_{i=0}^t \alpha_{i}$, it can be deduced that:
\begin{eqnarray}
q\left(\vx_t \mid \vx_0\right)  =  N\left(\vx_t ; \sqrt{\Bar{\alpha}_t} \vx_0,\left(1-\Bar{\alpha}_t\right) \mathbf{I}\right).
\label{eq:qxtx0}
\end{eqnarray}

Expanding above equation, we can express $\vx_t$ as a linear combination of $\vx_0$ and a noise variable $\epsilon$:
\begin{align}
    \vx_t  =  \sqrt{\Bar{\alpha}_t} \vx_0+\sqrt{1-\Bar{\alpha}_t} \epsilon,  
         \label{eq:x0-xt}
\end{align}
where $ \epsilon \sim N(0, 1)$ is noise satisfying the Gaussian distribution.


The image generation process gradually removes noise from random noise image $\vx_T \sim N(0, 1)$ through a learnable Markov chain, and its transition probability is:
\begin{align}
p_{\theta}\left(\vx_{t-1} \mid \vx_t\right)= N \left(\vx_{t-1} ; \mu_{\theta}\left(\vx_t, t\right), \Sigma\left(\vx_t, t\right)\right), 
\label{eq:p_theta}
\end{align}
where  $\theta$ represents the learnable parameter. If the learned reverse transition probability $ p_\theta(\vx_{t-1} \mid \vx_t) $ is close to $ q(\vx_{t-1} \mid \vx_t) $, the generation process can be achieved. 

The diffusion model is trained by maximizing the likelihood function, $\underset{\theta}{\max}\log p_\theta(\vx_0)$, but $p_{\theta}\left(\vx_0\right)$ cannot be directly optimized. Thus, DDPM uses variational inference to optimize the usual variational bound on negative log-likelihood:
\begin{align}
\label{eq:Lvlb}
-\mathbb{E}_{\vx_0}  \log p_{\theta}\left(\vx_0\right) 
\leq  \mathbb{E}_{q(\vx_{0: T})}\left[\log \frac{q\left(\vx_{1: T} \mid \vx_0\right)}{p_{\theta}\left(\vx_{0: T}\right)}\right] 
=  L_{VLB},
\end{align}

If the diffusion process $q\left(\vx_{t-1} \mid \vx_t, \vx_0\right)$ conditioned on $\vx_0$ and the generation process $p_{\theta}\left(\vx_{t-1} \mid \vx_t\right)$ are modeled as Gaussian with trainable mean functions and fixed variances, the objective in Eq. \ref{eq:Lvlb} can be simplified to: 
\begin{align}
L_{simple} = \mathbb{E}_{\vx_0, \epsilon}\left[ \left\|\epsilon-\epsilon_{\theta}\left(\sqrt{\Bar{\alpha}_t} \vx_0+\sqrt{1-\Bar{\alpha}_t} \epsilon, t\right)\right\|^{2}\right], 
\label{eq:losssimple}
\end{align}
where $\epsilon_\theta(\cdot)$ is the predicted noise value in generation process and $\epsilon$ is the injected noise value during the diffusion process.

\begin{figure*}[htbp]
	\centering
        \includegraphics[width=0.75\linewidth]{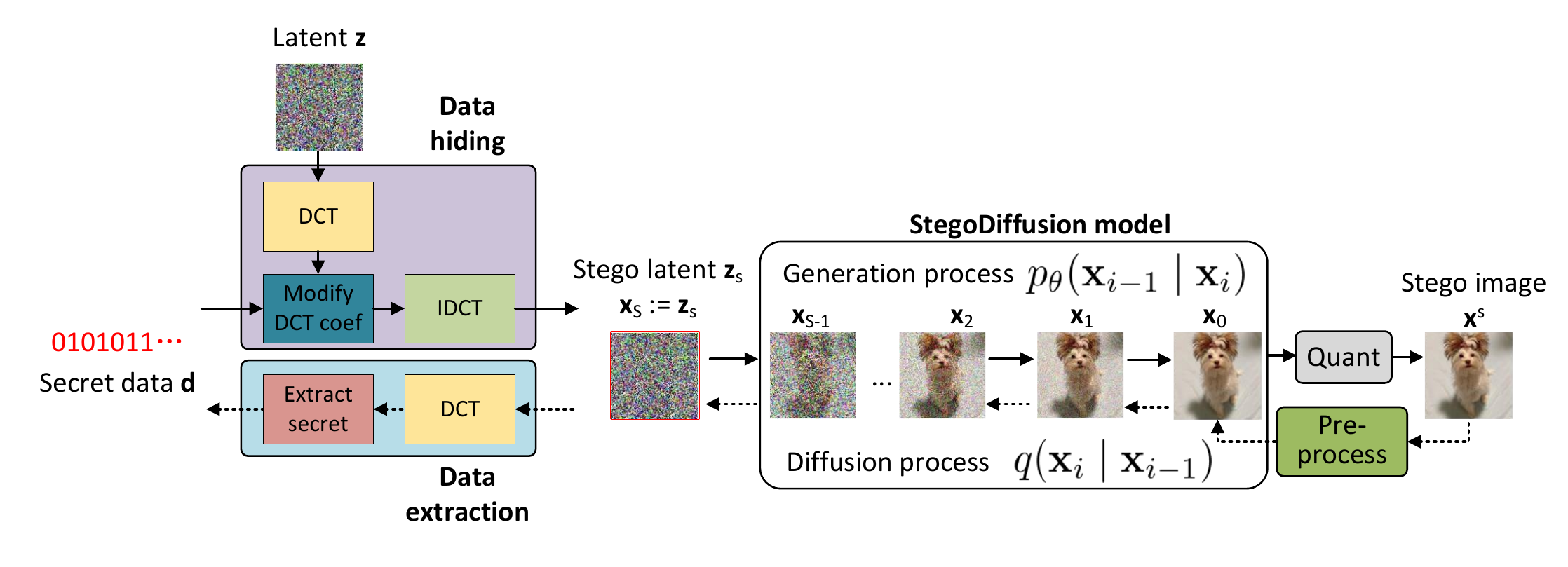}
	\caption{The flowchart of proposed GSD scheme for generative steganography.}
	\label{fig:overall}
\end{figure*}
\subsection{DDIM}
To sample more effectively, DDIM \cite{ddim} defines a non-Markovian chain to implement the forward diffusion process, as shown in Fig. \ref{fig:markov}b. The forward diffusion process can be derived from Bayes' theorem:
\begin{align}
q_{\sigma}\left(\vx_t \mid \vx_{t-1}, \vx_0\right)  = \frac{q_{\sigma}\left(\vx_{t-1} \mid \vx_t, \vx_0\right) q_{\sigma}\left(\vx_t \mid \vx_0\right)}{q_{\sigma}\left(\vx_{t-1} \mid \vx_0\right)}.
\end{align}
where the generation of $\vx_t$ depends simultaneously on $\vx_0$ and $\vx_{t-1}$ in non-Markovian diffusion processes. 

Given an input image $\vx_0$ and $\epsilon_t \sim N(0, 1)$, $\vx_t$ can be obtained through Eq. \ref{eq:x0-xt} in the forward diffusion process. After training, the neural network predicts the value of noise $\epsilon_t$ as $\epsilon_\theta(\vx_t, t)$. Therefore, the original input image $\vx_0$ can be predicted using $\vx_t$ and $\epsilon_\theta(\vx_t, t)$:
\begin{equation}
    \vx'_0 = \frac{\vx_{t}-\sqrt{1-\Bar{\alpha}_{t}} \epsilon_{\theta}\left(\vx_{t}, t\right)}{\sqrt{\Bar{\alpha}_{t}}}.
\end{equation}

The transition probability $p_\theta(\vx_{t-1} \mid \vx_{t})$ in the image generation process can be approximated by the transition probability $q_\sigma(\vx_{t-1} \mid \vx_t, \vx_0=\vx'_0)$ in the diffusion process. Therefore, $\vx_{t-1}$ can be obtained by sampling from $\vx_t$:
\begin{align}
\vx_{t-1}  =  &\sqrt{\Bar{\alpha}_{t-1}} \underbrace{\left(\frac{\vx_{t}-\sqrt{1-\Bar{\alpha}_{t}} \epsilon_{\theta}\left(\vx_{t}, t\right)}{\sqrt{\Bar{\alpha}_{t}}}\right)}_{\text {predicted } \vx_{0}} \nonumber \\
 & + \underbrace{\sqrt{1-\Bar{\alpha}_{t-1}-\sigma_{t}^{2}} \cdot \epsilon_{\theta}\left(\vx_t, t\right)}_{\text {direction pointing to \ } \vx_{t} }+\underbrace{\sigma_{t} \epsilon_{t}}_{\text {random noise}},
\label{eq:ddimsample}
\end{align}
where $\sigma_{t}=\eta \sqrt{(1-\Bar{\alpha}_{t-1}) / (1-\Bar{\alpha}_{t})} \sqrt{1-\Bar{\alpha}_{t} / \Bar{\alpha}_{t-1}}$, $\eta \in [0, 1]$. $\epsilon_{t} \sim N(0, 1)$, and $\epsilon_{\theta}(\vx_t, t)$ is the predicted value for $\epsilon_t$.

DDIM still uses a variational lower bound $J_{\sigma}$ to approximate the likelihood function of the image distribution:
\begin{equation}
-\mathbb{E}_{\vx_0}  \log p_{\theta}\left(\vx_0\right)= J_{\sigma} = \gamma \cdot L_{simple} + C,
\end{equation}
where $\gamma \in R^+$ and $C \in R$, $L_{simple}$ is the simplified loss function of DDPM in Eq. \ref{eq:losssimple}. Therefore, the variational lower bound $J_\sigma$ in DDIM can use the same optimization objective as in DDPM, and the training process is the same.

\section{Proposed Scheme}

This paper proposes a novel GS scheme called GSD, in which a reversible diffusion model (named StegoDiffusion) is developed for reversible stego image synthesis. StegoDiffusion adopts a deterministic non-Markovian chain with a fast sampling technique to speed up image generation. In this scheme, stego images and secret data are converted to each other through Euler iterations. Additionally, the DCT/IDCT transformations are employed to reduce the impact of image quantization and improve data extraction accuracy.

\begin{algorithm}[htb]
    \caption{Training strategy of StegoDiffusion in GSD.}
    \label{al:train}
    \begin{algorithmic}[1]
        \State \textbf{Input}: Hyperparameters $\Bar{\alpha}_t$ ($t \in [1, T]$);  neural network U-net; number of sampling steps $T$.
        \For{each step}
        \State Randomly sample input image $\vx_0 \sim q(\vx_0)$;
        \State Add random noise to $\vx_0$, $\vx_0 := \vx_0 + \vn$, where $\vn \sim N(0, 0.01^2)$;
        \State Randomly sample diffusion step $t \sim U(1, T)$;
        \State Randomly sample injected noise $\epsilon \sim N(0, 1)$;
        \State Compute the loss $L_{simple}$ and update U-net through gradients:
        $\nabla_{\theta}\left\|\epsilon-\epsilon_{\theta}\left(\sqrt{\Bar{\alpha}_t} \vx_0+\sqrt{1-\Bar{\alpha}_t} \epsilon, t\right)\right\|^{2}$.
        \EndFor
        \State \textbf{Output}: The trained StegoDiffusion model.
    \end{algorithmic}
\end{algorithm}

Fig. \ref{fig:overall} shows the overall framework of the proposed GSD, which mainly includes two important parts: hiding/extraction of secret data and the pre-trained StegoDiffusion model. During stego image generation, the secret data $\vd$ is first hidden in the DCT coefficients of the latent $\vz$, and then the stego latent $\vz_s$ is obtained through IDCT transformation. Next, $\vz_s$ is used as the initial noisy image $\vx_S$, and then the fast sampling technique is used to generate a stego image $\vx_0$, reducing the sampling step length of image generation from the original $T=1000$ to $S$ ($S$ much smaller than $T$), where the sampling image sequence is: $\vx_S \rightarrow \vx_{S-1} \cdots \rightarrow \vx_2 \rightarrow \vx_1 \rightarrow \vx_0$. After quantization (converting to quantized images with the pixel values in [0, 255]), a final stego image $\vx^s$ can be generated. When extracting hidden data, the pre-processed $\vx^s$ (converting quantized stego images into floating point images with the pixel values ranging in [-1, 1]) is input into StegoDiffusion, which gradually restores the initial noisy image through the diffusion process, $\vx^s \rightarrow \vx_{0} \rightarrow \vx_1 \cdots \rightarrow \vx_{S-1} \rightarrow \vx_S$. Then $\vx_S$ is used as the restored stego latent $\vz'_s$, and the hidden data $\vd'$ is extracted from the DCT coefficients of $\vz'_s$.

The StegoDiffusion in our scheme adopts a similar training strategy to DDPM. The main difference is that Gaussian noise is added to $\vx_0$ during the image pre-processing stage in our work to ensure data extraction accuracy. The training strategy of StegoDiffusion is shown in Algorithm \ref{al:train}. The hyperparameters $\Bar{\alpha}_t$ must be determined first. At each step, an image $\vx_0$, a time step $t$, and injected noise $\epsilon$ are randomly sampled. The neural network U-net outputs $\epsilon_\theta(\vx_t, t)$ as the predicted value for the injected noise $\epsilon$ at this moment. Then, the parameters of the U-net are updated to make the predicted noise $\epsilon_\theta(\vx_t, t)$ close to the injected noise $\epsilon$, where $\vx_t$ is replaced by $\sqrt{\Bar{\alpha}_t} \vx_0+\sqrt{1-\Bar{\alpha}_t} \epsilon$ following Eq. \ref{eq:x0-xt}.

\subsection{Reversibility of StegoDiffusion}
\label{sec:rever}
The generation and diffusion processes of StegoDiffusion are devised to be reversible for accurate data extraction from generated stego images. In Eq. \ref{eq:ddimsample}, the variance of the added noise is $\sigma_{t}=\eta \sqrt{(1-\Bar{\alpha}_{t-1}) /(1-\Bar{\alpha}_{t})} \sqrt{1-\Bar{\alpha}_{t} / \Bar{\alpha}_{t-1}}$, where $\eta \in [0, 1]$. When $\eta=0$, the added noise is zero and the sampling of the generation process becomes a deterministic process.
\begin{equation}
\vx_{t-1} = \frac{\vx_t}{\sqrt{\alpha_t}} + \left(  \sqrt{1-\Bar{\alpha}_{t-1}} - \sqrt{\frac{1- \Bar{\alpha}_{t}}{\alpha_t}} \right) \epsilon_\theta(\vx_t, t),
\label{eq:determinsample}
\end{equation}
where $\vx_{t-1}$ can be generated from $\vx_t$. When the number of iterations is $T$ and $\vx_T \sim N(0, 1)$, a series of content-deterministic images can be generated, $\vx_T \rightarrow \cdots \vx_t \rightarrow \vx_{t-1} \cdots \rightarrow \vx_0$.

By rearranging the above Eq. \ref{eq:determinsample}, the transition probability $p_\theta(\vx_{t-1} \mid \vx_{t})$ during the image generation process is:
\begin{equation}
     \frac{\mathbf{x}_{t-1}}{\sqrt{\Bar{\alpha}_{t-1}}}=\frac{\mathbf{x}_{t}}{\sqrt{\Bar{\alpha}_{t}}}+\left(\sqrt{\frac{1-\Bar{\alpha}_{t-1}}{\Bar{\alpha}_{t-1}}}-\sqrt{\frac{1- \Bar{\alpha}_{t}}{\Bar{\alpha}_{t}}}\right) \epsilon_{\theta}\left(\mathbf{x}_{t}, t\right),
     \label{eq:gen}
 \end{equation}
 let $\Delta t = 1$, the Eq. \ref{eq:gen} can be written as:
  \begin{equation}
     \frac{\mathbf{x}_{t-\Delta t}}{\sqrt{\Bar{\alpha}_{t-\Delta t}}} - \frac{\mathbf{x}_{t}}{\sqrt{\Bar{\alpha}_{t}}} = \left(\sqrt{\frac{1-\Bar{\alpha}_{t-\Delta t}}{\Bar{\alpha}_{t-\Delta t}}}-\sqrt{\frac{1- \Bar{\alpha}_{t}}{\Bar{\alpha}_{t}}}\right) \epsilon_{\theta}\left(\mathbf{x}_{t}, t\right),
     \label{eq:Euler}
 \end{equation}
  let $\sigma=(\sqrt{1-\bar{\alpha}} / \sqrt{\bar{\alpha}})$ and $\bar{\vx}= \vx / \sqrt{\bar{\alpha}}$ (both $\sigma$ and $\bar{\vx}$ are functions of $t$),  then Eq. \ref{eq:Euler} can be seen as an ordinary differential equation (ODE) solved by the Euler method:
 \begin{equation}
    \mathrm{d} \Bar{\vx}(t)= \epsilon_\theta\left(\frac{\Bar{\vx}(t)}{\sqrt{\sigma^{2}+1}}, t\right) \mathrm{d} \sigma(t),
\end{equation}
where $\Bar{\vx}_{t+1} = \Bar{\vx}_t + \mathrm{d} \Bar{\vx}(t) $ and $\Bar{\vx}_{t-1} = \Bar{\vx}_t - \mathrm{d} \Bar{\vx}(t) $. Then, according to $\vx_{t} = \sqrt{\Bar{\alpha}} \Bar{\vx}_{t} $, the image $\vx_t$ at any time $t$ can be solved using ODE solver. This means the StegoDiffusion model used in  the proposed GSD scheme is reversible.

If set $\Delta = -1$ in Eq. \ref{eq:Euler}, we can obtain the transition probability $q(\vx_{t+1} \mid \vx_t)$ for the diffusion process:
\begin{align}
    \frac{\mathbf{x}_{t+1}}{\sqrt{\Bar{\alpha}_{t+1}}} - \frac{\mathbf{x}_{t}}{\sqrt{\Bar{\alpha}_{t}}} = \left(\sqrt{\frac{1-\Bar{\alpha}_{t+1}}{\Bar{\alpha}_{t+1}}}-\sqrt{\frac{1- \Bar{\alpha}_{t}}{\Bar{\alpha}_{t}}}\right) \epsilon_{\theta}\left(\mathbf{x}_{t}, t\right), \nonumber  \\
     \mb{\vx}_{t+1} = \sqrt{\alpha_t} \vx_t +  \left(\sqrt{1-\Bar{\alpha}_{t+1}}- \sqrt{\alpha_t - \Bar{\alpha}_{t+1}}\right) \epsilon_{\theta}\left(\mathbf{x}_{t}, t\right).
     \label{eq:ext}
 \end{align}

In summary, the generation and diffusion processes of StegoDiffusion are a discrete form of the ODE. When the number of iterations is $T$ and $\vx_T \sim N(0, 1)$, the noisy image $\vx_T$ can be denoised step-by-step to obtain the image $\vx_0$ via Eq. \ref{eq:gen}, which is the image generation process. On the other hand, given $\vx_0$, the image $\vx_T$ can be restored step-by-step via Eq. \ref{eq:ext}, which is the diffusion process. These two processes are reversible, $\vx_T \leftrightarrow \vx_{T-1} \cdots \leftrightarrow \vx_1 \leftrightarrow \vx_0$.

\subsection{Stego Image Generation}
To resist the distortion of image quantization, we hide secret data in the frequency domain. First, hide secret data in the DCT coefficients of the latent $\vz$, where latent $\vz$ is a random noise that follows a standard Gaussian distribution, $\vz \sim N(0, 1)$.
\begin{align}
     \mb{Coeff_m} &= 1.0 \odot  \operatorname{reshape}  (2 \times \vd - 1),  \nonumber \\
      \vz_s &= \operatorname{IDCT} (\mb{Coef_m}),
\end{align}
where $\mb{Coeff_m}$ represents the DCT coefficients of the latent $\vz$ after hiding the secret data; $\odot$ denotes element-wise multiplication. The operator $\operatorname{reshape}(\cdot)$ reshapes $\vd$ into the same shape as the DCT coefficients of $\vz$. $\vd$ is the input binary secret data. When the secret data "1" needs to be hidden, the corresponding DCT coefficient of $\vz$ is modified to "+1.0"; when the secret data "0" needs to be hidden, the corresponding DCT coefficient of $\vz$ is modified to "-1.0". After embedding secret data in the DCT coefficients, the stego latent $\vz_s$ can be obtained by IDCT transformation.

Using $\vz_s$ as the initial noisy image $\vx_S$, and through the image generation process $p_\theta(\vx_{t-1} \mid \vx_t)$ of StegoDiffusion, the initial noisy image can be gradually denoised to generate a series of clear images, $ \vx_S := \vz_s \rightarrow \vx_{S-1} \cdots \vx_2 \rightarrow \vx_1 \rightarrow \vx_0$. After quantization, a stego image $\vx^s$ is obtained. The quantization operation converts the floating-point image $\vx_0$ into a digital image $\vx^s$ with pixel values in the range of [0, 255].

The maximum number of iteration steps $T$ in StegoDiffusion is set to a large value ($T=1000$) to ensure that the final noise image $\vx_T$ is close to a standard Gaussian distribution, i.e., $\vx_T \sim N(0,1)$. However, this also means that the image generation process requires step-by-step sampling for $T$ steps. The lengthy sampling process slows down the image generation speed, taking several seconds or even tens of seconds to generate each stego image, which is unfavorable for covert communication.

\begin{figure}[htb]
	\centering
	\includegraphics[width=0.8 \linewidth]{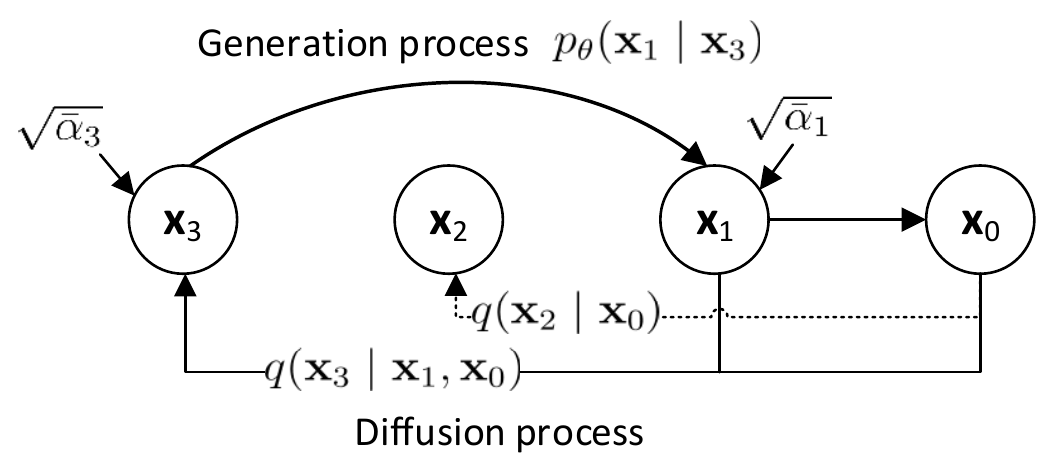}
	\caption{Illustration of fast sampling in StegoDiffusion. The sampling steps $\tau = \{t_1, t_2\} = \{1, 3\}$ are shown. }
	\label{fig:fast-sample}
\end{figure}

It is possible to speed up image generation by reducing the length of the sampling steps. Suppose that $\tau=\{t_1, t_2, \cdots, t_{i-1}, t_i, \cdots, t_S\}$ is a subsequence of the original sampling sequence $\mathcal{T}=\{1,2,\cdots,T\}$, where $\tau \subseteq \mathcal{T}$. By reducing the number of steps from $T$ to $S$ in the accelerated image generation process, the image generation speed is significantly increased. For example, if the original step sequence is $\{1, 2,\cdots, 1000\}$, and the accelerated sampling step sequence is $\{t_1=100, t_2=200, \cdots, t_{i-1}=100\times(i-1), t_i=100\times i, \cdots, t_{10}=1000\}$, the time required for image generation is only one-hundredth of the original time. The accelerated image generation process is illustrated in Fig. \ref{fig:fast-sample}. 
In the figure, the diffusion process ignores $\vx_2$ and generates $\vx_3$ from $\vx_0$ and $\vx_1$, and the transition probability in the diffusion process becomes $q(\vx_3 \mid \vx_1, \vx_0)$. During image generation, the transition probability for the generation process is $p_\theta(\vx_1 \mid \vx_3)$, and $\vx_1$ can also be obtained directly from $\vx_3$, skipping $\vx_2$. It should be noted that during StegoDiffusion training, each step $t \in [1, T]$ still needs to be considered, as shown in Algorithm \ref{al:train}. Only after the model is trained, several intermediate nodes are omitted to speed up the sampling process. The hyperparameter $\sqrt{\Bar{\alpha}_{t_i}}$ of the sampling node is determined by its index value $t_i$. In Fig. \ref{fig:fast-sample}, $\tau = \{t_1, t_2\} = \{1, 3\}$, the node $t_1 = 1$, thus $\sqrt{\Bar{\alpha}_{t_1}} = \sqrt{\Bar{\alpha}_1}$; and the node $t_2 = 3$, so the hyperparameter $\sqrt{\Bar{\alpha}_{t_2}} = \sqrt{\Bar{\alpha}_3}$. According to Eq. \ref{eq:ddimsample}, for any image sampling sequence $\tau = \{t_1, t_2, \cdots, t_{i-1}, t_i, \cdots, t_S\}$, the generation process is:

\begin{align}
\vx_{t_{i-1}}  = & \sqrt{\Bar{\alpha}_{t_{i-1}}}\left(\frac{\vx_{t_{i}}-\sqrt{1-\Bar{\alpha}_{t_{i}}} \epsilon_{\theta}\left(\vx_{t_{i}}, t_{i}\right)}{\sqrt{\Bar{\alpha}_{t_{i}}}}\right) \nonumber \\
&~~~+\sqrt{1-\Bar{\alpha}_{t_{i-1}}-\sigma_{t_{i}}^{2}} \cdot \epsilon_{\theta}\left(\vx_{t_{i}}, t_{i}\right)+\sigma_{t_{i}} \epsilon,
\end{align}
where $\sigma_{t_i} = \eta \sqrt{(1-\Bar{\alpha}_{t_{i-1}}) /(1-\Bar{\alpha}_{t_{i}})} \sqrt{1-\Bar{\alpha}_{t_{i}} / \Bar{\alpha}_{t_{i-1}}}$, and $\eta \in [0,1]$. When $\eta = 0$, the added noise is zero. At this time, image generation becomes a deterministic process.
Its transition probability $p_\theta(\vx_{t_{i-1}} \mid \vx_{t_i})$  can be expressed as:
\begin{align}
\vx_{t_{i-1}}  = \sqrt{\Bar{\alpha}_{t_{i-1}}}\left(\frac{\vx_{t_{i}}-\sqrt{1-\Bar{\alpha}_{t_{i}}} \epsilon_{\theta}\left(\vx_{t_{i}}, t_{i}\right)}{\sqrt{\Bar{\alpha}_{t_{i}}}}\right)
+\sqrt{1-\Bar{\alpha}_{t_{i-1}}} \cdot \epsilon_{\theta}\left(\vx_{t_{i}}, t_{i}\right).
\label{eq:fastsample}
\end{align}

The process of generating a stego image is shown in Algorithm \ref{al:gen}. First, the input secret data is embedded into the latent $\vz$ to obtain the stego latent $\vz_s$. Then, using the transition probability in Eq. \ref{eq:fastsample}, the stego image can be generated step-by-step by denoising: $\vx_{t_S} \rightarrow \vx_{t_{S-1}} \dots \rightarrow \vx_{t_2} \rightarrow \vx_{t_1} \rightarrow \vx_0$. The sampling sequence $\tau = \{t_1, t_2, \cdots, t_{i-1}, t_i, \cdots, t_S \} \subseteq \{1, 2, \cdots, T\}$. For convenience, $\vx_{t_S} \rightarrow \vx_{t_{S-1}}  \dots \rightarrow \vx_{t_2} \rightarrow \vx_{t_1} \rightarrow \vx_0$ can be renumbered as $\vx_{S} \rightarrow \vx_{S-1}  \dots \rightarrow \vx_{2} \rightarrow \vx_{1} \rightarrow \vx_{0}$, and the transition probability $p_\theta(\vx_{t_{i-1}} \mid \vx_{t_i})$ is simplified as $p_\theta(\vx_{i-1} \mid \vx_i)$, as shown in Algorithm \ref{al:gen}. For example, when $\tau = \{100, 200, \cdots, 900, 1000\}$ and $S=10$, it means that only ten nodes subscripted by these numbers are selected in image generation. The corresponding stego images $\vx_{10} \rightarrow \vx_{9} \rightarrow \vx_{8}  \dots \rightarrow \vx_{1} \rightarrow \vx_{0}$ are generated as the result. Then $\vx_{0}$ is quantized to obtain the stego image $\vx^s$.

\begin{algorithm}[htb]
\caption{Stego Image Generation}
\label{al:gen}
\begin{algorithmic}[1]
\State \textbf{Input:} Pre-trained StegoDiffusion; Preset hyperparameters $\Bar{\alpha}_t$ ($t \in [1, T]$); Sampling sequence $\{t_1, \cdots, t_S \} \subseteq \{1, 2, \cdots, T\} $.
\For{each step}
\State Input the secret data $\vd \in \{0,1\}^n$;
\State Randomly sample latent $\vz \sim N(0, 1)$;
\State Hide $\vd$ into the latent $\vz$ to obtain the stego latent $\vz_s$;
\State Use $\vz_s$ as the initial noise image $\vx_S $, and input it into the pre-trained StegoDiffusion;
\State Iterate the generation process $p_\theta(\vx_{i-1} \mid \vx_i)$ to generate stego image, $\vx_S \cdots \rightarrow \vx_1 \rightarrow \vx_0 $;
\State Quantize $\vx_0$ to obtain the final stego image $\vx^s$.
\EndFor
\State \textbf{Output:} The set of stego images.
\end{algorithmic}
\end{algorithm}

\subsection{Secret Data Extraction}

\begin{algorithm}[htb]
\caption{Secret Data Extraction}
\label{al:ext}
\begin{algorithmic}[1]
\State \textbf{Input}: Pre-trained StegoDiffusion model same with the sender, preset hyperparameters $\Bar{\alpha}_t$ ($t \in [1, T]$), a sampling sequence $\tau = \{t_1, \cdots, t_S \} \subseteq \{1, 2, \cdots, T\}$, and the received stego images.
\For{each step}
\State Input a stego image $\vx^s$;
\State Pre-process $\vx^s$ and convert it to a floating-point image $\vx_0$;
\State Use the diffusion process $q(\vx_{i+1} \mid \vx_i)$ to recover the original image sequence $ \vx_0 \rightarrow \vx_1 \cdots \rightarrow \vx_S $;
\State Use $\vx_S$ as the recovered stego latent $\vz'_s$.
\State Extract the secret data $\vd'$ from the DCT coefficients of $\vz'_s$.
\EndFor
\State \textbf{Output}: The extracted secret data.
\end{algorithmic}
\end{algorithm}

\begin{align}
    \label{eq:dataext}
    & \mb{Coeff_r}  = \operatorname{DCT}(\vz'_s), \nonumber  \\
    & \vd'  =  \operatorname{reshape} \left( \left\lceil \frac{\operatorname{Sign} (\mb{Coeff_r}) + 1}{2} \right\rceil \right),
\end{align}
where $\mb{Coeff_r}$ is the DCT coefficients containing secret data, $\operatorname{Sign}(\cdot)$ output sign matrices of input, and $\lceil \cdot \rceil$ denotes the upward rounding operation. $\operatorname{reshape}(\cdot)$ operation reshapes the output $\vd'$ to match the same shape of the original secret data $\vd$.

Upon receiving the stego image $\vx^s$, the receiver can extract the secret data using the diffusion process of StegoDiffusion, shown as Algorithm \ref{al:ext}.

According to Eq. \ref{eq:fastsample} and the reversibility of StegoDiffusion described in section \ref{sec:rever}, the transition probability of the accelerated diffusion process $q(\vx_{t_{i+1}} \mid \vx_{t_i})$ is:
\begin{align}
\vx_{t_{i+1}}  &= \sqrt{\Bar{\alpha}_{t_{i+1}}}\left(\frac{\vx_{t_{i}}-\sqrt{1-\Bar{\alpha}_{t_{i}}} \epsilon_{\theta}\left(\vx_{t_{i}}, t_{i}\right)}{\sqrt{\Bar{\alpha}_{t_{i}}}}\right) 
+\sqrt{1-\Bar{\alpha}_{t_{i+1}}} \cdot \epsilon_{\theta}\left(\vx_{t_{i}}, t_{i}\right).
\label{eq:fastdiffuse}
\end{align}

When extracting hidden secret data, the starting image is the stego image $\vx^s$, and by iterating through the diffusion process in Eq. \ref{eq:fastdiffuse}, the noise image can be gradually restored, i.e., $\vx^s \rightarrow \vx_{t_1} \cdots \rightarrow \vx_{t_{i}} \rightarrow \vx_{t_{i+1}} \rightarrow \vx_{t_{S-1}} \rightarrow \vx_{t_S}$. For convenience, they are renumbered as follows: $\vx^s \rightarrow \vx_{1} \cdots \rightarrow \vx_{i} \rightarrow \vx_{i+1}  \cdots \rightarrow \vx_S := \vz'_s$. And  $q(\vx_{t_{i+1}} \mid \vx_{t_i})$ is  simplified as $q(\vx_{i+1} \mid \vx_{i})$.
The noise image $\vx_S$ is the restored latent stego $\vz'_s$, and hidden secret data can be extracted according to Eq. \ref{eq:dataext}.

\section{Experimental Results}

\begin{figure*}[htb]
\centering
\includegraphics[width=0.85\linewidth, trim=0 49 0 0, clip]{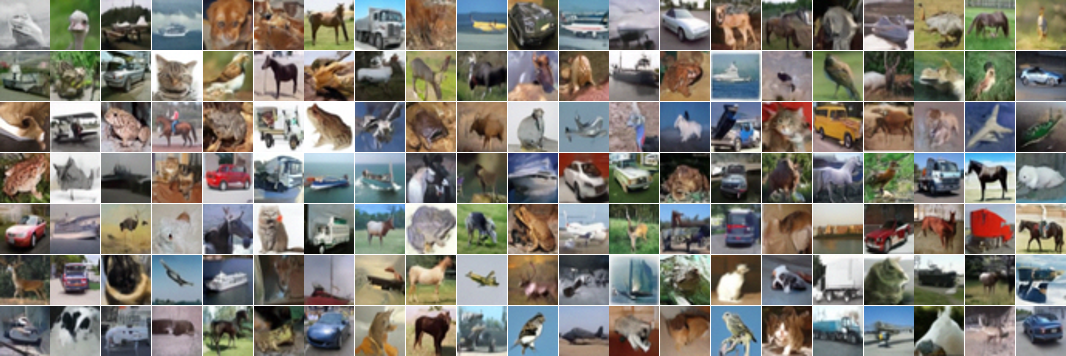}
\caption{Generated $32\times32$ stego landscape images on Cifar10 dataset. The payload=1bpp, Acc=99.91\%, Pe=0.4574, Fid=7.57,  \textit{S}=50.}
\label{fig:cifar}
\end{figure*}

\begin{figure*}[htbp]
\centering
\begin{minipage}{\linewidth}
\centering
\subfloat[Stego face images on CelebA dataset]{\includegraphics[width=0.85\linewidth, trim=0 97 0 0, clip]{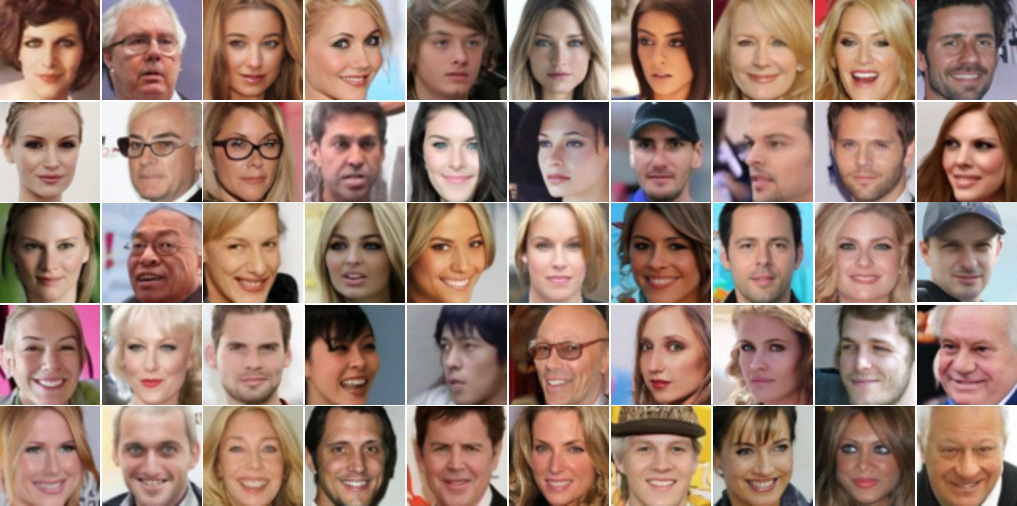}
\label{fig:face} }

\end{minipage}
\begin{minipage}{\linewidth}
\centering

\subfloat[Stego bedroom images on Bedroom dataset]{
\includegraphics[width=0.85\linewidth, trim=0 97 0 0, clip]{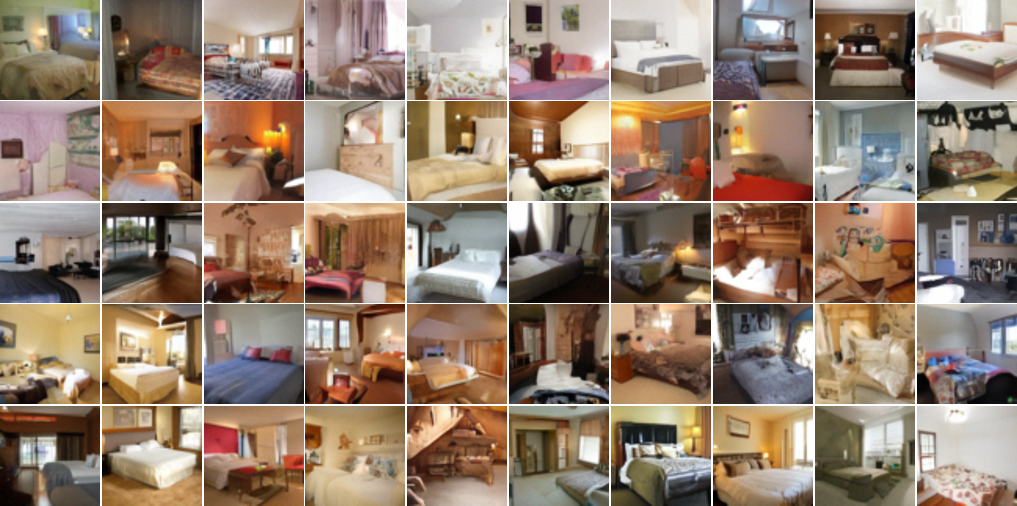}
\label{fig:bedroom} }

\end{minipage}
\caption{Generated $64 \times 64$ steg images by GSD on different datasets. For CelebA dataset: Payload=1 bpp, Acc=100.00\%, Pe=0.4850, Fid=8.10, and \textit{S}=50; For Bedroom dataset: Payload=1 bpp, Acc=100.00\%, Pe=0.4620, Fid=11.38, and \textit{S}=50.}
\label{fig:img}
\end{figure*}

\subsection{Implemented Details}
The experiments are conducted on Ubuntu 18.04, using an Nvidia 3090 graphic card and the Pytorch 1.8 platform. Adam optimizer with a default learning rate of 4e-2 is used for model training. During training, the diffusion step $T = 1000$ and the hyperparameters are set as $\alpha_t = 1- \frac{0.02t}{T}$ ($t \in [1, T]$), and $\Bar{\alpha}_t = \prod_{i=1}^{T}\alpha_i$.  The process of accelerated image generation requires $S$ sampling steps to generate the stego image, where $\tau = \{t_1, t_2, \cdots, t_{i-1}, t_i, \cdots, t_S \}$, and $\tau \subseteq \{1, 2, \cdots, T\}$. For the $S$ sampling steps, a uniform sampling interval is used, $\Delta t = t_i - t_{i-1} = \frac{T}{S}$. For example, when $T=1000$ and $S=100$, the sampling interval is $\Delta t = \frac{1000}{100}=10$, and the sampling step sequence $\tau = \{ 10, 20, 30 \cdots 990, 1000\}$. In the experiments, different sampling steps $S$ are used, including 10, 50, 100, 200, 500, and 1000, with corresponding sampling intervals $\Delta t$ of 100, 20, 10, 5, 2, and 1, respectively. 

CelebA \cite{celeba}, Cifar10 \cite{cifar10}, and Lsun-Bedroom \cite{lsun} are used for model training. CelebA and Lsun-Bedroom were used to generate 64×64 stego images, while Cifar10 was used to generate 32×32 stego landscape images in the experiments. The model was trained on approximately 130 million images in each dataset.

\subsection{Metrics}
We use five evaluation metrics, bpp, Fid, Acc, Pe, and Time, to evaluate the performance of the GSD scheme, which respectively represent data hiding capacity (payload), image quality, extraction accuracy of secret data, security, and time consumption.

Metric bpp represents the number of hidden data bits per pixel (bit-per-pixel), $\operatorname{bpp} = \frac{\operatorname{len}(\vd)}{C \times H \times W}$, where $\operatorname{len}(\vd)$ represents the number of hidden data bits; $C$, $H$, and $W$ respectively represent the number of channels, height, and width of the stego image.

Fid is used to evaluate the quality of generated images, $\operatorname{Fid}  = \left\|\mu_{r}-\mu_{g}\right\|^{2}+\operatorname{Tr}\left(\Sigma_{r}+\Sigma_{g}-2\left(\Sigma_{r} \Sigma_{q}\right)^{1 / 2}\right)$, where $\mu_{r}$ and $\mu_{g}$ represent the mean of features extracted from real images and generated images, respectively; $\Sigma_{r}$ and $\Sigma_{g}$ represent the covariance of extracted features; $\operatorname{Tr}$ denotes the trace of a matrix. A smaller Fid value indicates better quality of generated images.

Acc represents the extraction accuracy of hidden secret data, $\mathrm{Acc}=\frac{\vd\odot \vd'}{\operatorname{len}(\vd)}$, where $\vd$ and $\mathbf{d'}$ respectively represent the hidden secret data and the extracted hidden data. $\odot$ denotes the element-wise Exclusive NOR (XNOR) operation.

Pe is the most commonly used evaluation metric for steganalysis security, $\mathrm{Pe}=\mathrm{min}_{P_{FA}}\frac{1}{2}(P_{FA} + P_{MD})$, where $P_{FA}$ and $P_{MD}$ are the false alarm rate and miss-detection rate in steganalysis. Pe ranges from 0 to 1, with an optimal value of 0.5 indicating the highest security level for steganographic algorithms.

Time represents the average time to generate the stego images or extract the hidden data, $ \operatorname{Time}  = \frac{t_{gen} + t_{ext}}{2}$, where $t_{gen}$ is the average time to generate stego images from secret data and $t_{ext}$ is the average time to extract hidden data from the stego images. The unit of Time is second/image.

\subsection{Effectiveness of GSD}

\begin{table}[htb]
    \centering
    \small
   \caption{The comprehensive performance of the proposed GSD scheme on different datasets}
    \label{tab:perf}
    \setlength{\tabcolsep}{2.5 mm}
    \begin{tabular}{l c c c c c} \toprule
        Dataset & $S$ & Acc$(\%)\uparrow$ & Pe$\rightarrow0.5$ & Fid$\downarrow$ & Time$\downarrow$ \\ \midrule
        \multirow{4}{*}{} & 10 & 99.66 & 0.4552 & 15.34 & 7.33e-3  \\
         \multirow{4}{*}{\begin{tabular}[c]{@{}c@{}}Cifar10 \\ 32×32\end{tabular}} & 50 & 99.91 & 0.4574 & 7.57 & 3.62e-2  \\
         \multirow{4}{*}{ } & 100 & 99.93 & 0.4595 & 7.45 & 0.0861 \\ 
         \multirow{4}{*}{ } & 200 & 99.93 & 0.4664 & 8.08 & 0.1605 \\ 
         \multirow{4}{*}{ } & 500 & 99.94 & 0.4796 & 9.10 & 0.3509 \\ 
         \multirow{4}{*}{ } & 1000 & 99.97 & 0.4857 & 13.55  & 0.7710 \\ \midrule
         \multirow{4}{*}{ } & 10 & 99.84 & 0.4743 & 15.77  & 4.02e-2  \\ 
         \multirow{4}{*}{ \begin{tabular}[c]{@{}c@{}}CelebA \\ 64×64\end{tabular}} & 50 & 100.00 & 0.4850 & 8.10  & 0.1346  \\ 
         \multirow{4}{*}{ } & 100 & 100.00 & 0.4575 & 9.28 & 0.2520 \\ 
         \multirow{4}{*}{ } & 200 & 100.00 & 0.4967 & 11.61 & 0.4893 \\ 
         \multirow{4}{*}{ } & 500 & 100.00 & 0.4781 & 13.91 & 1.2071 \\ \midrule
        \multirow{4}{*}{ } & 10 & 99.51 & 0.4530 & 17.24 & 1.78e-2  \\ 
         \multirow{3}{*}{\begin{tabular}[c]{@{}c@{}}Bedroom \\ 64×64\end{tabular} } & 50 & 99.93 & 0.4620 & 11.38  & 0.1094  \\ 
         \multirow{4}{*}{ } & 100 & 99.98 & 0.4698 & 12.46   & 0.2128 \\ 
         \multirow{4}{*}{ } & 200 & 99.99 & 0.4647  & 14.20 & 0.4684 \\ 
         \multirow{4}{*}{ } & 500 & 100.00 & 0.4730 & 16.17  & 0.9954  \\ 
         \bottomrule
    \end{tabular}
\end{table}

The GSD scheme is trained on Cifar10 \cite{cifar10}, CelebA \cite{celeba}, and Lsun-bedroom \cite{lsun}, and can generate stego images of size $32 \times 32$ and $64 \times 64$ with a payload of 1 bpp, as shown in Figs. \ref{fig:cifar} and \ref{fig:img}. 

Fig. \ref{fig:cifar} illustrates 10 classes of randomly generated  stego landscape images of size $32 \times 32$, each of which appears to be clear and realistic. Fig. \ref{fig:face} shows stego face images of size $64 \times 64$, with reasonable facial features and natural appearances. Fig. \ref{fig:bedroom} shows stego bedroom images of size $64 \times 64$, which are almost indistinguishable from real ones. High-quality stego images can deceive the third party and make it difficult to detect any anomalies in the images, thus increasing the security of covert communication.

The comprehensive performance of the proposed GSD scheme is evaluated at different sampling steps $S$ on different datasets, shown in Table \ref{tab:perf}. On different datasets, the secret data payload for generated stego images of different sizes is consistently 1 bpp, which means that larger images can hide more secret data. As the image generation sampling step $S$ increases, the values of the Acc, Pe, and Time metrics gradually increase, while the Fid values first decrease and then gradually increase. When $S$ is between 50 and 100, all indicators are balanced: the image generation is fast, Acc can reach around 100\%, the Pe values tested by SCRMQ1 \cite{color_rich} are between 0.45 and 0.5, and Fid achieves smaller values. At this time, excellent performance is achieved in terms of image generation speed, extraction accuracy, image quality, and security.

\begin{figure}[htb]
	\centering
	\includegraphics[width=0.9\linewidth]{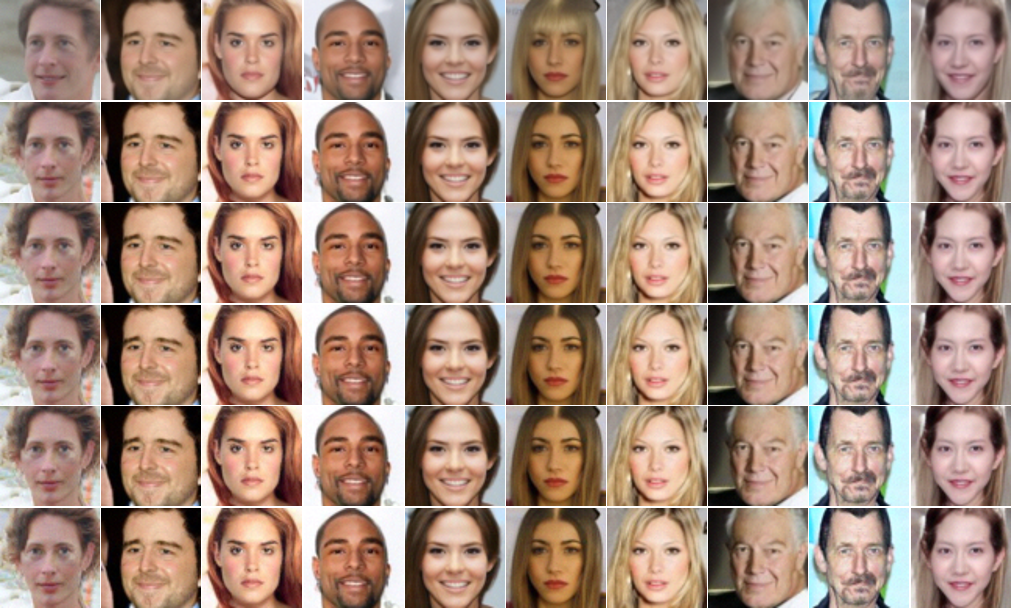}
	\caption{Generated stego face images by GSD with different sampling steps \textit{S}. From the first row to the last row, \textit{S} used in each row is 10, 50, 100, 200, 500, and 1000 respectively.}
	\label{fig:celeba}
\end{figure}

To study the influence of the sampling steps $S$ on the content of the generated stego images, some stego images are generated using the same settings but varying $S$ on the CelebA dataset, as presented in Fig.\ref{fig:celeba}. It can be observed that the content of the generated stego images is extremely similar using different $S$. Only the images generated with $S=10$ (the top row) have slight differences compared to the others. When $S=50$, the generated image content is similar to that generated with larger $S$, but the image generation speed is much faster.

\subsection{Visualizing Reversibility}
\begin{figure}[ht]
\centering
\includegraphics[width=0.95\linewidth]{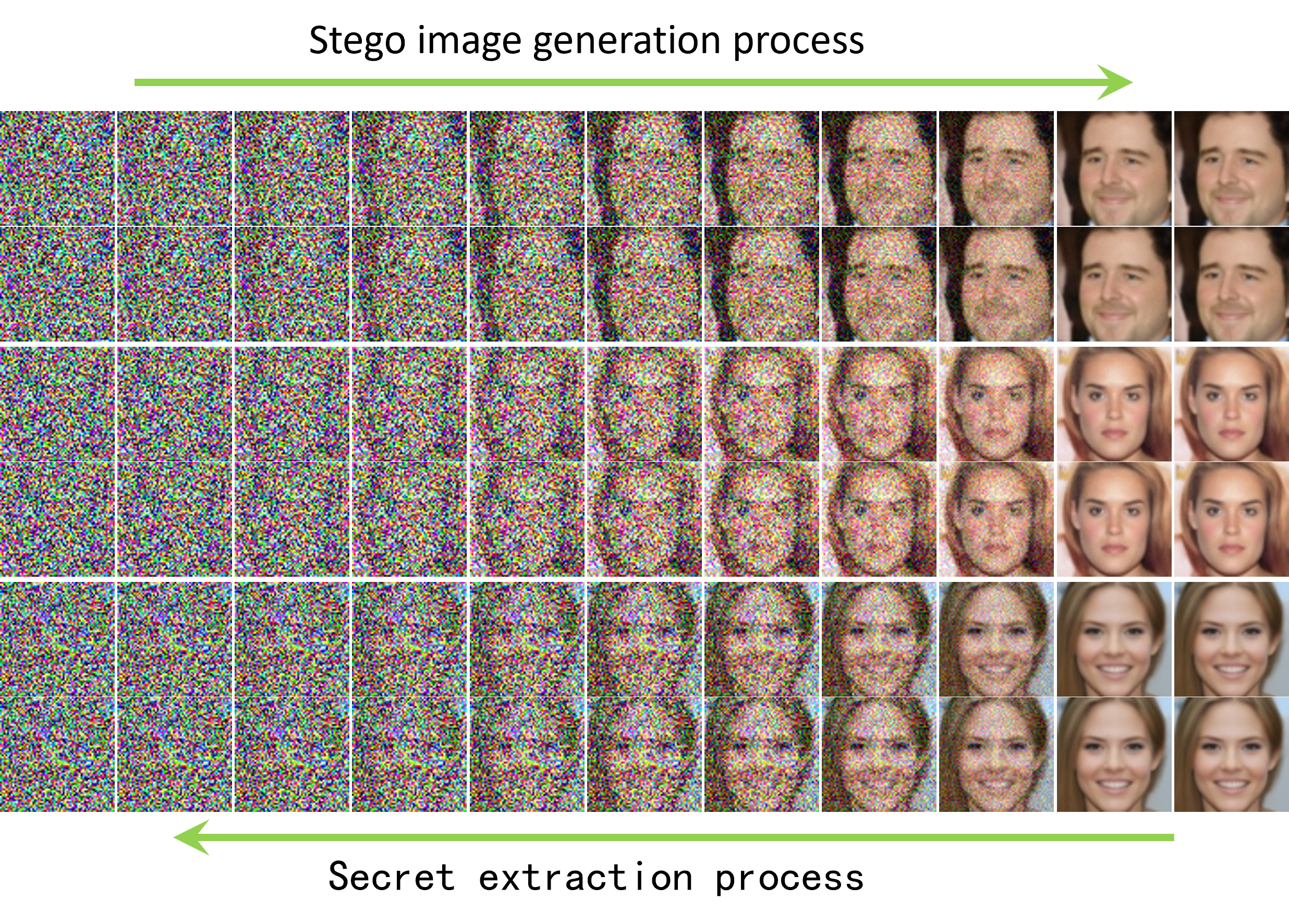}
\caption{The process of step-wise stego image generation and secret data extraction in GSD. The first, third, and fifth rows show the gradual denoising process to generate a stego image from the stego latent $\vz_s$, while the second, fourth, and sixth rows show the stepwise restoration process of $\vz_s$ from the quantized stego image.}
\label{fig:inv}
\end{figure}

To visualize the reversibility of the GSD scheme, we visually present the process of generating a stego image and extracting hidden secret data in Fig. \ref{fig:inv}, where the image generation step $S=10$. From left to right, the images in the 1st, 3rd, and 5th rows show the process of generating a stego image from the stego latent $\vz_s$, using the generation process of proposed StegoDiffusion. A sequence of denoised stego images is generated according to Eq.\ref{eq:fastsample}, $ \vx_{10} := \vz_s \rightarrow \vx_9 \rightarrow \vx_8 \cdots \rightarrow \vx_1 \rightarrow \vx_0 $.  
From right to left, the images in the 2nd, 4th, and 6th rows show the process of gradually recovering the stego latent from the quantized stego image, through the diffusion process of StegoDiffusion as described in Eq.\ref{eq:fastdiffuse}, $\vx_0 \rightarrow \vx_1 \cdots \rightarrow \vx_9 \rightarrow \vz'_s := \vx_{10} $. 
From Fig. \ref{fig:inv}, it can be seen that the noise in the stego latent $\vz_s$ are gradually eliminated through the generation process, and the image becomes clearer. After 10 iterations, a realistic and natural stego image can be obtained. When the generated stego image undergoes the diffusion process, the removed noise can be gradually restored.  
The absolute mean error of the input stego latent $\vz_s$ and the recovered stego latent $\vz'_s$ is less than 1e-2. This indicates that the image generation process and the secret data extraction process are almost reversible, and the secret data hidden in the stego latent can be extracted with an accuracy close to or equal to 100\%. 
When using larger step $S$, higher stego image quality and data extraction accuracy can be obtained.

\subsection{Performance Comparison}

We compare the proposed GSD with other advanced methods, as shown in Table \ref{tab:comp}. Methods Hu \cite{hu2018novel}, Liu \cite{ideas}, and GSN \cite{gsn} all generate stego images using generative adversarial networks (GANs). GSF \cite{gsf} is a GS scheme based on the Flow model. GSD is our proposed GS scheme based on the proposed StegoDiffusion model, with a sampling step $S=50$ during image generation.

\begin{table}[htb]
	\centering
        \small
        \setlength{\tabcolsep}{3 mm}
	\caption{Comparison of our GSD with SOTA methods}
	\label{tab:comp}
	\begin{tabular}{ccccc}
		\toprule
		Method                        & bpp $\uparrow$         & Acc (\%) $\uparrow$   & Fid $\downarrow$   & Pe $\rightarrow 0.5 $ \\ 
		\midrule
        Liu \cite{ideas}					 & 4.17e-2	& 98.26		& 26.37	& 0.5350			\\
        Hu \cite{hu2018novel}                 & 2.44e-2    & 91.73      & 30.81     & 0.4700          \\ 
        GSN \cite{gsn}                         & 0.33  & 98.15 & 12.88  & 0.4756      \\
	GSF \cite{gsf}                        & 1   & 99.96   
        & 22.45  & 0.4864   \\
  		
           \textbf{Proposed GSD}             & \textbf{1} & \textbf{100.00} & \textbf{8.10}  & \textbf{0.4960}   \\ 
    \bottomrule

	\end{tabular}
\end{table}

Comparative experiments are conducted on the CelebA \cite{celeba} dataset to generate $64\times 64$ sized stego images. When measuring the Fid, 50,000 generated and real images are used. Pe is obtained through the steganalysis network Ye-net \cite{yenet}, which is trained on 10,000 generated stego/cover images and tested on 1,000 stego/cover images, where the cover image is generated directly by latent $\vz$ without hiding secret data.
There are trade-offs among different metrics, such as improving payload may result in decreases in the other metrics. Liu \cite{ideas} and Hu \cite{hu2018novel} reduced secret data payload to improve extraction accuracy and image quality, which is reflected in bpp lower than 1e-2 and Fid around 30. Compared to Liu \cite{ideas} and Hu \cite{hu2018novel}, GSN \cite{gsn} and GSF \cite{gsf} improve the payload and image quality while maintaining comparable extraction accuracy and security. However, the four schemes mentioned above cannot achieve 100\% extraction accuracy. Our GSD outperforms the other methods in all metrics: the payload is up to 1 bpp, and the extraction accuracy can reach 100\% while retaining  the best image quality and security.


\subsection{Security Analysis}

\begin{figure}[htb]
	\centering
	\begin{minipage}{0.49\linewidth}
		\centering
		\includegraphics[width=\linewidth]{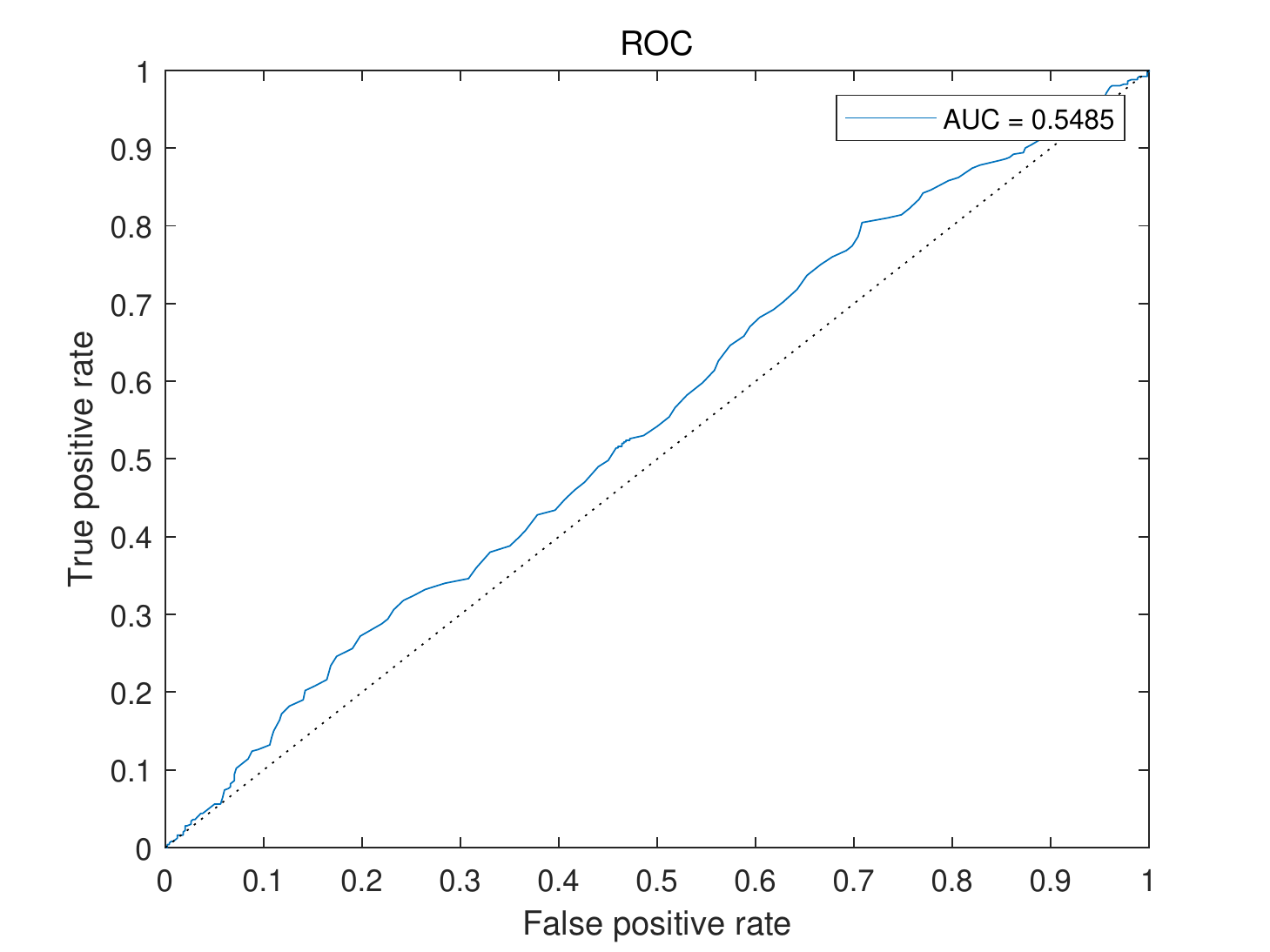}
		\subcaption{CelebA}
		\label{roc-celeba}
	\end{minipage}
	\begin{minipage}{0.49\linewidth}
		\centering
		\includegraphics[width=\linewidth]{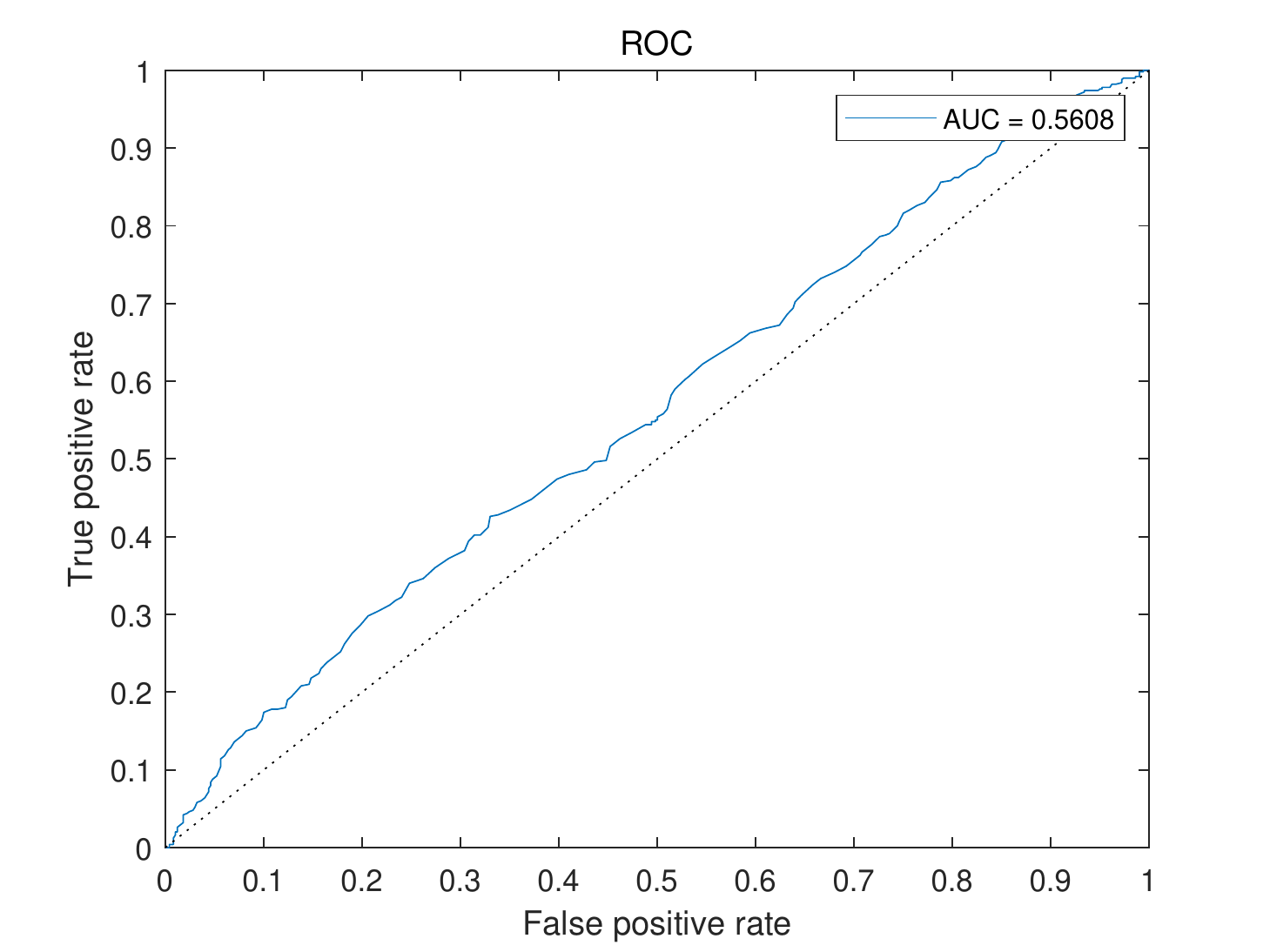}
		\subcaption{Bedroom}
		\label{roc-bedroom}
	\end{minipage}
	\caption{Detection ROC curves on different datasets.}
	\label{fig:roc}
\end{figure}

To verify the security of the proposed GSD scheme, we test the resistance of GSD to the steganalysis algorithm SCRMQ1 \cite{color_rich} on datasets CelebA and Bedroom. The sampling step $S$ is set to 50 when generating stego/cover images. In the test, 1,000 stego images containing secrets and 1,000 generated cover images without secrets are classified, and then the ROC curves for detection are plotted in Fig. \ref{fig:roc}. In steganography, the optimal ROC curve is the counter-diagonal (represented by the dashed line in the figure), and the optimal AUC value is 0.5. As shown, our ROC curves and AUC values are close to the ideal results, which suggests that our proposed GSD scheme is highly secure.

\subsection{Conclusion}

This paper proposes a high-performance generative steganography scheme named "Generative Steganography Diffusion" (GSD) by devising an invertible StegoDiffusion model. The GSD scheme leverages a non-Markov diffusion process with a fast sampling technique to speed up image generation. By constructing an ordinary differential equation based on the transition probability in StegoDiffusion, the scheme enables the inverse transformation of the stego image and secret data through the Euler iterative method. The experimental results demonstrate the superiority of GSD over existing methods across all evaluation metrics.

\bibliographystyle{ACM-Reference-Format}
\bibliography{ref}

\end{document}